# Humans, Machine Learning, and Language Models in Union: A Cognitive Study on Table Unionability

Sreeram Marimuthu*
Nina Klimenkova*
smarimuthu@wpi.edu
nklimenkova@wpi.edu
Worcester Polytechnic Institute
Worcester, Massachusetts, USA

Roee Shraga
rshraga@wpi.edu
Worcester Polytechnic Institute
Worcester, Massachusetts, USA

## Abstract
Data discovery and table unionability in particular became key tasks in modern Data Science. However, the human perspective for these tasks is still under-explored. Thus, this research investigates the human behavior in determining table unionability within data discovery. We have designed an experimental survey and conducted a comprehensive analysis, in which we assess human decision-making for table unionability. We use the observations from the analysis to develop a machine learning framework to boost the (raw) performance of humans. Furthermore, we perform a preliminary study on how LLM performance is compared to humans indicating that it is typically better to consider a combination of both. We believe that this work lays the foundations for developing future Human-in-the-Loop systems for efficient data discovery.

## 1 Introduction

Data discovery is a fundamental challenge in contemporary data science, enabling users to explore, understand, and expand their datasets by uncovering valuable new data sources [5, 16]. In this context, the table union task plays a pivotal role by identifying candidate tables that can potentially extend existing datasets with additional instances [14]. However, the definition of table unionability itself presents a challenge.

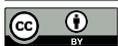

**Figure 1: Example Questions from the Main Section of the Survey illustrating two view, namely, the unionability questions itself (1) and confidence assessment and explanation (2).**

Traditional definitions state that unionable tables should share compatible schemas [17], but practical implementations often relax this constraint. Nargesian et al. [14] consider tables unionable if their attributes are from the same domain, while Khatiwada et al. [12] further require that unionable tables share the same relationships between columns. These varying definitions create ambiguity about what constitutes a "correct" judgment of unionability. The top part of Figure 1 illustrates this ambiguity with two sample tables of a question from our experiment. While both tables contain information about geographical locations and languages, they have different structures and partially overlapping content. Determining whether these tables should be considered unionable involves complex semantic judgments about attribute correspondence and data compatibility. We argue that determining whether tables can be unionable can benefit from human judgment, especially when dealing with heterogeneous data sources. Thus, in this study we take *a closer look at how humans perceive and understand unionability*.

Human intelligence has traditionally been considered irreplaceable in data discovery and integration tasks. For example, Li's examination of human-in-the-loop data integration [13] highlights how, despite advances in automated techniques, humans possess unique abilities to understand nuances and interpret ambiguous situations that algorithms struggle with. This human advantage becomes particularly evident in tasks requiring semantic understanding and contextual knowledge [4] , such as determining table unionability. Yet, our understanding of how humans actually make these judgments and where their cognitive limitations might affect the discovery process remains limited.

The cognitive aspects of human judgment in data integration have been explored by Ackerman et al. [1], who revealed systematic biases affecting human matching decisions. Their cognitive model identified patterns of overconfidence, where human confidence in determining correspondences often exceeds actual accuracy, creating a measurable gap between perceived and actual performance. Their research also found that decision time correlates with both confidence and accuracy, with longer deliberation typically associated with more difficult judgments and lower confidence. Understanding these cognitive patterns becomes essential when evaluating whether tables from diverse sources can be meaningfully combined. In our work, we extend this cognitive approach to the specific task of table union judgment, examining how human decision patterns, confidence levels, and response times influence unionability assessments.

Finally, it is important to note that with the emergence of Large Language Models (LLMs) like GPT, the landscape of data discovery is evolving. These models offer promising capabilities for data discovery tasks, including table union decisions, yet they too have limitations. Recent work by Brown et al. [2] has demonstrated that LLMs can perform complex reasoning tasks through few-shot learning and in-context examples. However, LLMs still struggle with tasks requiring domain-specific knowledge or precise semantic understanding without sufficient context [21] . We investigate the potential of large language models to perform table union decisions both in isolation and enhanced with human input through in-context learning, we will discuss it in more detail in Section 5.

**Contributions:** We present an experimental study examining human decision-making in table union scenarios and explore how these insights can enhance automated approaches, improve labeling quality, and inform benchmarks for evaluating table union methods. The main takeaway is that by *understanding human input patterns, we can significantly enhance the quality of table unionability judgments*, improve automated systems, and potentially provide robust benchmarks for future research. Our main contributions are:

- A comprehensive behavioral analysis of human judgment in table union tasks, highlighting cognitive patterns, biases, and significant confidence-accuracy gaps through metrics such as confidence levels, response times, and interaction patterns.
- A machine learning framework, assessing the reliability of human unionability judgments, leveraging behavioral indicators to improve the quality and enhance human-labeled data.
- A preliminary investigation comparing human, LLM, and collective intelligence for table union decisions, demonstrating the effectiveness of combining human and AI capabilities and identifying promising directions for future human-AI collaborations.

**Related Work:** Despite extensive research on human involvement in data discovery and integration tasks, systematic evaluation of human judgment in table union search remains largely unexplored. Human input has traditionally been studied in related areas such as schema matching and entity resolution. For instance, crowdsourcing and pay-as-you-go frameworks have frequently relied on human validators to confirm algorithmic outputs [7, 13, 23] . However, these approaches typically assume human judgments to be accurate or treat them as ground truth without systematically investigating their reliability or potential biases.

Recent studies have focused on human-in-the-loop approaches for feature and dataset discovery, exploring the challenges users face when searching for relevant data [9–11]. Hulsebos et al. [9] highlighted ongoing difficulties users encounter during dataset search processes, particularly emphasizing usability and cognitive challenges. Similarly, Ionescu et al. [10, 11] investigated human-in-the-loop methods for feature discovery in tabular data, providing valuable insights into user behavior and interactive exploration, though without specifically addressing unionability judgments.

Specifically for table union search, automated techniques have leveraged contextual representations and contrastive learning methods. Fan et al. [6] proposed capturing column contexts to improve accuracy in union search tasks, yet their method focused primarily on algorithmic performance without examining how humans approach unionability decisions. Hu et al. [8] introduced Auto-TUS, employing tabular representation learning for automatic table union search, yet without a systematic analysis of human judgment patterns. Our work aims to fill the gap by systematically evaluating human decision-making processes and identifying underlying cognitive patterns and biases.

Finally, related research also investigated human behavior in data matching scenarios through a metacognitive lens [1, 20, 22]. Shraga and Gal [20] identified patterns of overconfidence and correlations between decision times and accuracy. Subsequent efforts further characterized human matching experts based on behavioral indicators [19] and introduced frameworks such as HumanAL to calibrate human judgment across diverse matching tasks [18]. Our current study extends these findings specifically to the context of table union search.

## 2 Studying Humans in Table Unionability

We designed an experimental user study to investigate human perception of table unionability with a particular focus on understanding behavioral factors such as confidence level, and decision time and Human accuracy for each question. The experiment was structured into three stages and conducted through an online Qualtrics[1] survey. We set up a semi controlled environment to collect data on how participants evaluate the unionability of tables. Significantly, by construction, the formal task we are analyzing here is table union *classification* (rather than *search*) [15]. In future research we may also investigate human search rather than judgment.

Our participant pool consisted exclusively of students from our institution, including undergraduate, graduate, and PhD students from disciplines such as Data Science, Computer Science, Artificial Intelligence, and related fields. All participants had prior knowledge of data structures and relational data models, with varying levels of experience in data integration and discovery tasks. This helped them to understand schema overlaps and improved the reliability of their unionability assessments.

---
[1]https://www.qualtrics.com/

To ensure data quality and avoid any potential biases, each participant could only take the survey once and only through their institutional email. We used the Qualtrics' built in authentication and tracking features to enforce this. This also enabled us to collect useful metadata such as start date, end date, progress, browser, operating system, and human-interaction patterns including the recorded timestamps of first click, last click, and page submit, along with the total number of clicks made during that question. More similar information about the survey is described in detail in the following subsection.

## 2.1 Survey Design and Deployment

Participants were authenticated before starting the experimental survey, restricting them access to a single response. The first section of the survey introduces the experiment to all the participants, outlining that determining table unionability requires evaluation across multiple factors and that interpretations may vary across different contexts. We also asked participants to provide demographic information such as age, education level, major, and English proficiency to understand our participant pool. Importantly, institutional emails were only used for authentication and were not retained, ensuring participant anonymity.

In the next section of the survey (i.e., the main experiment), participants were randomly and evenly assigned to one of four survey versions (V1–V4), each comprising of 8 table unionability assessment questions. This balanced design enabled consistent structure while supporting comparative analysis across different versions. Each question was split across two pages, as shown in Figure 1, the first one recorded participants' unionability judgment and background behavioral factors such as decision time and click patterns associated with their decision; the second page collected self-reported confidence level of their decision (via a mandatory slider initialized at 50) and optionally asked them for an explanation. The final section, concluded with a common reflective questionnaire to all the participants with multiple definitions of unionability applied to a fixed set of tables.

For convenience, we created an open version of our survey[2].

## 2.2 Dataset Curation and Feature Engineering

To establish high-quality data, only fully completed responses were considered for our analysis. This filtering mitigated any biases due to incomplete submissions. In total, the dataset comprised of 58 authenticated participants, each contributing 8 unique judgments on table unionability questions, totaling 464 human-annotated responses. The tables used for unionability assessments were adapted from the UGEN benchmarks introduced by Pal et al. [15]. We chose 13 table pairs representing both unionable (with 1 to 4 common columns) and non-unionable configurations. These were extracted from publicly available benchmark ground truth files[3]. Each pair was presented as a simplified subset of the full benchmark tables into two versions. One version kept the original column headers, while another version removed those headers, which gave us 26 table pair combinations in total. To design the 4 versions of the main experiment (each with 8 table questions), we drew 16 unionable and 16 non-unionable pairs from this pool. The pairs were chosen so that every version had an equal level of difficulty; some pairs appeared twice, others once, and a few were excluded, as a result. Participants were evenly distributed across the survey versions, but when only considering completed responses, they were split across with V1 (27.6%), V2 (22.4%), V3 (25.9%), and V4 (24.1%). Age-wise, the majority (65.5%) fall into the 18-24 range, with the rest between 25-34. Educationally, 50% are master's, 31% are PhD., and the rest 19% are bachelor's students. English proficiency was generally high, with more than 70% identifying as fluent or native speakers. Technically, the cohort was also well qualified, with 81.1% majoring in Data and Computer Science disciplines.

After explicitly cleaning the exported raw data file from Qualtrics, we initially had 21 attributes, capturing participant metadata, demographics, behavioral indicators (e.g., click counts, confidence, decision time, explanations), and question-specific attributes. This was later extended to 36 attributes through various types of normalization, label encoding, derived metrics, and feature engineering, including both human-annotated labels and ground truth labels from prior benchmarks to capture participant behavior and data discovery task complexity.

## 3 Observations and Insights

We now present a detailed analysis of the responses and how accuracy varied across different categories, highlighting key trends and differences observed in the dataset. A large proportion of the responses (83.4%) included written explanations for their decision, suggesting that the participants were generally thoughtful. Interestingly, most submitted answers (65%) use a single-click, indicating fast or intuitive decisions while 21% employed 2 or 3 clicks, and the remaining subset ranges from 4 to up to 46 clicks. These numbers show how varied the clicking patterns are from different participants, suggesting various decision strategies.

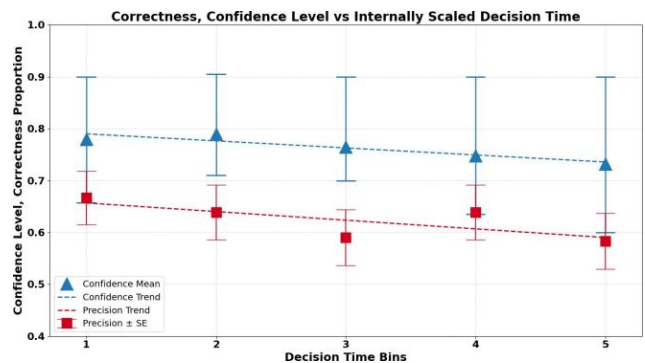

Figure 2: Correctness and Confidence vs Scaled Decision Time

Figure 2 illustrates how the correctness proportion (red-dotted line) and the mean confidence level (blue-dotted line) vary with binned decision times. We first discarded outlier participants who took 100 minutes or longer to complete the survey. Then we removed any data lying beyond ± 2 units of standard deviation from the mean within that subset. The remaining decision times were scaled internally[4] (individually), for each user, by linearly transforming their decision times associated with the table unionability tasks to a 0–1 range. This allowed us to uniformly assess decision times across users and divided them into 5 equally sized bins.

---

[2] https://wpi.qualtrics.com/jfe/form/SV_8jqmFQVl43NcHci
[3] https://github.com/northeastern-datalab/gen/blob/main/data

> **Example 3.1.**
> *For example, say user A has decision times (in seconds)* [28, 126, 138, 256]. *After internally scaling we will get* [0, 0.429, 0.482, 1]. *Also consider User B with* [16, 30, 6, 61] *which are scaled to* [0.178, 0.435, 0, 1].
> *Here, while the users have very different highest decision time values (256 & 61), the scaling considers both as 1, indicating that this is the highest decision time for the user.*

For correctness, Figure 2 plots the mean of each bin with standard errors as vertical ranges, and for confidence, we plot the mean values with interquartile ranges (IQR) [3]. Notably, confidence trends slightly downward from left to right bins (0.79 to 0.74), interestingly each bin's 75$^{th}$ percentile remains around 0.90. Meanwhile, the precision tracks roughly in parallel, dropping from 0.66 to 0.59. Individually each bin shows minor deviations, but overall both lines follow similar directions, suggesting at a slight decrease in accuracy and confidence as the decision time goes up.

| User | Ans. | Explanation |
|---|---|---|
| 1_1 | Yes | Both tables have 5 rows and 3 columns, meaning that both tables have 5 records and 3 attributes. Also, there's only one data type present in each table (string/text). Given that both tables have the same number of attributes and the same data type, both tables can be unioned at both the row level and the column level. |
| 1_11 | No | If all the column names would have been same, it would be easy for making the union operation easier. But Table B consists additional information than what is there in Table A. Hence, I think that for one to make a union table between them, there would be some computations & changes made which makes it difficult. That's reason I think that it is No. |

**Table 1: Examples of User Explanations for Question shown in Fig. 1**

> **Example 3.2.**
> *Recall Figure 1, presenting two unionable tables containing geographical information (continents, countries, languages, cities). Table 1 provides explanations from two participants who offered detailed reasoning, yet reached opposite conclusions. User 1_1 viewed the tables as structurally similar (3 columns, all string-based data) and judged them as unionable. In contrast, User 1_11 pointed to mismatched column names and the extra information in Table B, concluding that unionability was questionable. Though both responses are logical, only User 1_1 answered the question correctly with respect to the ground truth.*
> *When examining behavioral and cognitive factors, notable differences arise: User 1_1 submitted with a single click in 28.03 seconds and reported 91% confidence. Meanwhile, User 1_11 changed their response once (2 clicks), took 102.79 seconds, and reported 70% confidence. Despite divergent reasoning and behavior, both illustrate how users apply different criteria to the unionability task.*

Examining the performance across the survey versions revealed that V1 has the highest overall accuracy (70.3%), while V2 and V3 were both around 57.5–57.7%, and V4 was close to 58.9%. Such differences may come from variations in question difficulty, phrasing, or interface presentation in each version.

While taking into account the English proficiency levels, those indicating "Native Speaker" ability exhibited a higher average accuracy (63.0%) compared to "Fluent" (60.4%) and "Proficient" (62.5%). The group at "Intermediate" level scored the lowest (55.0%), this outcome is rather expected, as subtle language differences could contribute to misunderstandings of the questions, thereby reducing accuracy. Notably, not a single participant identified themselves at "Basic" (least ranked) level.

When considering click counts, single-click submissions ($n$ = 300) attained 62.0% accuracy. Those who clicked 2 ($n$ = 72) or 3 ($n$ = 26) times achieved 58.3% and 73.1%, respectively, while 4-click respondents ($n$ = 15) dropped to 33.3%. Notably 5-click submissions ($n$ = 16) from 9 different participants rose sharply to 93.8%. Beyond raw click counts, responses with written explanations (83.4% of the data) showed a moderate accuracy advantage (62% vs. 58%) over those without. In sum, the interplay between number of clicks, explanations open up diverse approaches to the decision tasks, with varying levels of success depending on the individual respondent. This leads us to the next section, aiming to utilize these key factors to improve our human-annotated labels.

## 4 Calibrating Human Table Unionability Labels

We set out to investigate whether we could utilize the behavioral and cognitive factors of participants, influencing their table unionability assessment to effectively improve upon our human-annotated labels. We employ traditional machine learning (ML) models for this study, over different feature subsets derived from participant interactions and responses. Each survey version served as its own test set, with the remaining three versions comprise collectively as the training set. We trained four ML models namely, *Logistic Regression (LR)*, *k-Nearest Neighbors (KNN)*, *Random Forest (RF)*, *XGBoost (XGB)* as binary classifiers and selected the best-performing one on the test set. Note that for training we are using "ground truth" labels curated by us.

Our motivation was to determine which of these factors hold the most predictive power for determining a correct response, and how they compare to the baseline accuracy achieved entirely by human labels and a default ML model with all the features. While designing our experiments, we began with 33 features that spanned across click-based behaviors, demographics, metadata, human labels (including explanations), decision times, and confidence levels. Firstly, we look into Table 2, which shows a comparison of accuracy for human responses and a default ML model across all 4 survey versions. We use accuracy (encoding Yes = 1 and No = 0) as in our context 'Yes' and 'No' predictions are similarly important. As we see across the table, the ML model consistently outperformed human responses in terms of accuracy. On average, the ML model achieved an accuracy of 0.77, +25.5% over human baseline with 0.61.

---
[4]within responses of the same user (individually) as explained in Example 3.1

This gap was particularly notable in V3, where ML accuracy reached 0.88 compared to 0.58 for humans, a significant increase of 52.2%. The complete classification report which we did not include here due to space constraints, confirmed that ML also distinguished the classes much better than humans, supported by higher F1-scores for both classes (0 and 1) in all the versions as shown in Table 2, hinting not only higher accuracy but also more consistent predictions. The results support the potential of integrating Human labels and associated features with traditional ML systems to improve the reliability in similar data discovery tasks.

| Survey | Human | | | ML (All Features) | | | Best Model |
|---|---|---|---|---|---|---|---|
| | Acc. | F1\|0 | F1\|1 | Acc. | F1\|0 | F1\|1 | |
| V1 | 0.70 | 0.72 | 0.69 | 0.83 (+17.8%) | 0.83 | 0.83 | LR |
| V2 | 0.58 | 0.57 | 0.59 | 0.64 (+10.1%) | 0.65 | 0.61 | KNN |
| V3 | 0.58 | 0.58 | 0.57 | 0.88 (+52.2%) | 0.86 | 0.89 | XGB |
| V4 | 0.59 | 0.57 | 0.60 | 0.73 (+24.2%) | 0.73 | 0.74 | RF |
| Avg. | 0.61 | 0.61 | 0.61 | 0.77 (+25.5%) | 0.77 | 0.77 | - |

Table 2: Accuracy raw Human labels vs. ML enhance labels

After training each ML model on *all* 33 features by default, we proceeded to isolate 6 distinct aggregated subsets based on their semantic similarities and data collection methodologies as shown in Table 3. This collective features approach was designed to pinpoint which set of factors contributed the most to predictive performance. All the features and detailed explanation of each subset, including individual attributes and their computation methods, are available in our online repository[5].

| Feature Group | Description |
|---|---|
| Click | Click behavior metrics |
| User | Demographics & metadata |
| Human-Labels | Participant response items |
| Quantified-Human-Labels | Group-level correctness |
| Decision-Time | Temporal decision measures |
| Confidence Level | Self-reported confidence |

Table 3: Aggregated Subsets with their descriptions.

Figure 3 zooms in on the accuracy of the model when using only the aforementioned subsets. *Quantified* features produced the largest improvement over the human baseline in V1, with a gain of 33.4% the highest overall accuracy across all combinations. In V2, *confidence* features led to the greatest increase (+11.6%), while in V4, *decision time* features displayed the highest gain (+48.5%). Interestingly, in V1, 2 subsets outperformed the default *all* features model, while V2 and V4 each had one cognitive subset accomplish the same. Notably in V3, three different subsets, *quantified*, *decision time*, and *all features* each improved accuracy by the exact same margin of 52.2%, indicating the successful predictive strength of certain subsets outweigh the *all features* set in most survey versions and only able to match the same in V3.

When observing the average performances, compared against Human responses in every survey version, notably, utilizing *all* 33 features yielded a substantial 25.5% improvement on an average over the human baseline across all survey versions. Moreover, focusing exclusively on subsets, *decision time* features often proved even more beneficial on average (+32.3%), emphasizing how temporal patterns can strongly indicate correctness and when only considering *quantified* features it showed (+20.1%) improvements. Conversely, only relying on *user* features typically underperformed (−18.1% on average), suggesting these attributes alone only offer limited insight into participant decision-making.

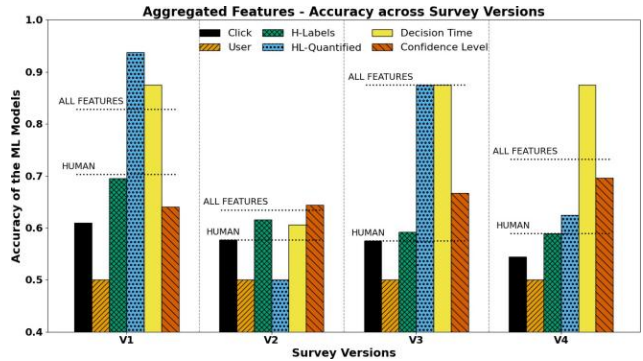

Figure 3: Comparison of Accuracy over Aggregated Subsets

Overall, these findings demonstrate that aggregated subsets such as decision times, quantified labels, and confidence levels can significantly improve human-annotated labels. While some groups performed really well, others were not as successful, combining them with more structured measures can show good improvements, reinforcing the idea that how individuals approach a task can be just as influential as the task itself. Finally, it is noteworthy that these behavioral and cognitive factors can be used to annotate or improve the existing benchmarks even better.

## 5 Do LLMs benefit from Human Inputs?

As a preliminary study, we focused on open-source models and among these we chose *Llama-3.3 70B Instruct* model.[6] Our main objective was to evaluate if incrementally adding human insights could progressively enhance the LLM's performance. We compared four scenarios, varying levels of human-derived context:

(1) **Human (Actual):** Our Baseline, average performance based on actual (raw) human responses from the Survey.
(2) **Human (Majority):** This shows how accuracy would shift if one simply follows the majority vote for each question.

---
[5] https://github.com/sreerammarimuthu/table-unionability-study

[6] https://huggingface.co/meta-llama/Llama-3.3-70B-Instruct

(3) **LLM:** Provided with only the table descriptions.
(4) **Human + LLM (Added Context):** Included numeric consensus from human responses with the average human meta-cognitive metrics such as decision time and confidence level (e.g., "..9 out of 13 humans answered yes,.. an average Decision Time of 61.742 seconds,.. an average Confidence score of 74.917/100..") for this particular question.

Each scenario was designed independently, with LLMs processing purely textual prompts without any retention of previous context, thus ensuring each prediction relied entirely on the immediate input provided. The results are presented in Table 4 for each survey version (V1-V4) for comparison.

| Survey | Human | | LLM | Human + LLM |
|---|---|---|---|---|
| | Actual | Majority | Actual | Added Context |
| V1 | 0.70 | **1.0 (+42.2%)** | 0.63 (-11.1%) | 0.75 (+6.7%) |
| V2 | 0.58 | 0.50 (-13.3%) | 0.50 (-13.4%) | **0.63 (+8.3%)** |
| V3 | 0.58 | **0.88 (+52.2%)** | 0.63 (+8.7%) | **0.88 (+52.2%)** |
| V4 | 0.59 | 0.63 (+6.1%) | 0.63 (+6.1%) | **0.75 (+27.3%)** |
| Average | 0.61 | **0.75 (+22.7%)** | 0.59 (-2.8%) | **0.75 (+22.7%)** |

**Table 4: Comparison of Accuracy: Human vs. LLMs. Percentages are relative to the Human-Actual column**

Overall, the results indicate a positive effect of adding human insights into the LLM's decision-making process. Nevertheless, in V1, *Human Majority* category alone achieves 100.0% accuracy, a remarkable +42.2% improvement over the *Human Actual* accuracy (our baseline) which was just 70.3%. Notably, it improved the baseline accuracy for V3 even more by +52.2% on par with the performance of the *Human+LLM* category.

Across other versions, the human baseline averages to about 61.1%, while relying solely on the *LLM* scenario alone achieved slightly lower accuracy 59.4% on average (−2.8%) confirming the fact that individually Humans performed better than LLMs. However, integrating additional human quantified consensus and meta-cognitive metrics such as decision time and confidence level in *Human+LLM* significantly enhanced the performance and recovered back to the accuracy of 75.0% (+22.7%). This suggests that additional context enables the model to make more accurate decisions. Specifically, the improvement observed in the *Human+LLM* scenario across different survey versions indicates that the additional human metrics provide essential context that helps the model interpret ambiguous or borderline cases more effectively. This is particularly clear in cases such as V3 and V4, where the combined *Human+LLM* accuracy matches or even exceeds *Human Majority* accuracy alone, showcasing the model's ability to leverage intricate human insights effectively.

Our experiments show that human inputs can significantly improve LLM performance on tasks involving refined human judgments such as table unionability. Note that this is not trivial as it might seem, e.g., by comparing this to in-context learning [2]. The raw results provided by humans, as we show in Sections 3-4, are far from perfect. Also note that both concise human input and detailed human input produced the best outcomes.

Finally, while LLMs benefited from human inputs, an additional (offline) analysis revealed that the LLM did not consistently improve through the addition of meta-cognitive factors alone. This suggests current LLMs may still lack refined capabilities to fully utilize such complex human-derived inputs. We intend to further analyze this in future research.

## 6 Limitations and Challenges

While this study provides insights into human decision-making in table union tasks, several limitations and challenges remain. Addressing these issues presents opportunities for future research to improve the reliability and applicability of table union judgments:

- **Exploring additional LLMs**: The current study utilized only the Llama 3.3 70B model; exploring and comparing other advanced LLMs such as GPT-4o could provide deeper insights and potentially better performance.
- **Hybrid approaches (ML + LLMs)**: Integrating traditional machine learning predictions into LLM inputs represents an unexplored direction that may further enhance the accuracy and reliability of table union judgments.
- **Participant diversity and scale**: Expanding the participant pool could yield more generalizable insights and allow more nuanced analysis of cognitive biases and behavioral patterns.
- **Comprehensive questionnaire analysis**: Further investigation of questionnaire data could reveal additional behavioral or demographic factors influencing decision-making.
- **Expanded ML algorithms and feature engineering**: While the current work focused solely on some of the traditional machine learning algorithms, as we mentioned in previous sections, future research could explore a broader set of machine learning algorithms and more extensive feature engineering approaches to better capture human judgment nuances.
- **Analysis of question-specific responses**: Examining how different formulations or definitions of unionability questions impact human responses could yield important insights into variability and reliability of judgments.

## 7 Conclusion

This study provides insights into human judgment behaviors and cognitive biases in the context of table union tasks. Our systematic behavioral analysis revealed key insights into cognitive patterns influencing human judgment, emphasizing the value of understanding these behaviors to enhance decision quality. Additionally, we introduced and evaluated machine learning and LLM-based frameworks, revealing that leveraging collective human intelligence enhances decision accuracy beyond individual human or standalone automated methods. Our findings emphasize the potential for effective human-AI collaboration in data discovery tasks, providing a foundation for future work aimed at refining and optimizing hybrid human-machine approaches.

## Acknowledgments

This work was supported in part by NSF under award number IIS-2325632. We acknowledge the mentorship of Gerardo Vitagliano.